\documentclass[aps,prl,preprint,groupedaddress]{revtex4}
\usepackage{mathrsfs,txfonts,graphicx}
\catcode`@=11
\@ifundefined{pdfoutput}{\usepackage[dvips,bookmarks]{hyperref}}{\usepackage{hyperref}}
\catcode`@=12
\hypersetup{colorlinks,bookmarksopen,bookmarksnumbered,citecolor=blue,pdfstartview=FitH}
\def\hhref#1{\href{http://arxiv.org/abs/hep-th/#1}{hep-th/#1}}
\def\nohhref#1{\href{http://arxiv.org/abs/#1}{#1}}
\def\phref#1{\href{http://arxiv.org/abs/hep-ph/#1}{hep-ph/#1}}
\def\mhref#1{\href{mailto:#1}{#1}}
\def\half{{\textstyle{1\over{\raise.1ex\hbox{$\scriptstyle{2}$}}}}}

\begin{document}
\preprint{YITP-SB-08-30}
\title
    {
        Worldgraph Approach to Yang-Mills Amplitudes\\ from N=2 Spinning Particle
    }
\author
    {
        Peng Dai\footnotemark \footnotetext{Electronic address: \mhref{pdai@grad.physics.sunysb.edu}},
        Yu-tin Huang\footnotemark \footnotetext{Electronic address: \mhref{yhuang@grad.physics.sunysb.edu}} and
        Warren Siegel\footnotemark \footnotetext{Electronic address: \mhref{siegel@insti.physics.sunysb.edu}}
    }
\affiliation
    {
        C. N. Yang Institute for Theoretical Physics, Stony Brook University, Stony Brook, NY 11790-3840
    }
\date{\today}
\begin{abstract}
    By coupling the N=2 spinning particle to background vector fields, we construct Yang-Mills amplitudes for trees and one loop. The vertex operators are derived through coupling the BRST charge; therefore background gauge invariance is manifest, and the Yang-Mills ghosts are automatically included in loop calculations by worldline ghosts. Inspired by string calculations, we extend the usual worldline approach to incorporate more ``generalized" 1D manifolds. This new approach should be useful for constructing higher-point and higher-loop amplitudes.
\end{abstract}
\maketitle

\section{I. Introduction}

First-quantization has provided an efficient way of calculating Yang-Mills amplitudes. A set of rules for writing down 1-loop Yang-Mills amplitudes was first derived by Bern and Kosower from evaluating heterotic string amplitudes in the infinite string-tension limit \cite{Bern:1991aq}. Later an alternative derivation of the same rules (but only for the 1-loop effective action) from first-quantization of the particle was given by Strassler \cite{Strassler:1992zr}. It is worth noting that Bern and Kosower's string approach naturally produces whole non-abelian amplitudes including both one-particle-irreducible (1PI) and non-1PI amplitudes since such a distinction was not present at the string level. On the other hand Strassler's first-quantized particle approach is based on a path integral representation of the effective action, and therefore is intrinsically more convenient for calculating the effective action itself, or 1PI amplitudes. In addition, the generalization of these first-quantized rules to multi-loop amplitudes has not been clear. A generalization of Bern and Kosower's approach seems difficult, because the calculation of multi-loop string amplitudes is very complicated. On the other hand, although there has been successful generalization of Strassler's approach to derive 2-loop Yang-Mills amplitudes from first-quantization of the particle with the sewing method \cite{Sato:1998sf}, a more concise representation for Yang-Mills amplitudes that is easier for generalization to all loop levels will be preferred, if it exists. In fact, such a representation has not yet been found even for Yang-Mills tree amplitudes. This is partially because the vacuum, ghost measure and Green function needed for the calculation of trees and multi-loops have not been clarified. Although there are already many ways to compute Yang-Mills tree amplitudes, it is important to clarify how first-quantization of the particle works at tree level for the purpose of generalizing this method to both non-1PI cases and multi-loop level. This is the main purpose of this paper.
 
To derive the first-quantized rules for trees, we start from theories of free relativistic spinning particles, which were first developed by Brink et al.\ \cite{Brink:1976uf} and many others \cite{Howe:1988ft}. In these theories the spin degree of freedom is encoded in the worldline supersymmetry. More precisely, the BRST quantization of the particle action with $N$-extended worldline supersymmetry shows that the cohomology is of a spin-$\frac{N}{2}$ particle.

In this paper we study the $N=2$ theory, which describes a spin-1 particle. We derive the vertex operator for background gauge fields via the usual BRST quantization method, thus ensuring background gauge invariance.  (The coupling of background vector fields to the spin-$\frac{1}{2}$ particle was formulated in \cite{Brink:1976uf}. It was used to calculate effective actions in \cite{Bastianelli:2002qw}.)  We proceed to show how the correct amplitudes can be derived. In the usual worldline approach, all interactions are derived by coupling external fields to the 1-dimensional worldline or loop. For higher-point tree graphs this approach usually requires sewing lower-point tree graphs to the worldline. This does not fit the picture of perturbation in the first-quantized theory since one would normally anticipate a formalism that can calculate any amplitude without calculating the lower-point amplitudes first: The knowledge of Green functions and vertex operators should be sufficient. Here we propose an alternative (``worldgraph") approach that includes 1D topological spaces that are not strictly 1D manifolds: They are not always locally $R^{1}$, but only fail to be so at a finite number of points. Taking these spaces into account we derive a set of rules for computing amplitudes that can be extended to all possible graphs.

We organize this paper as follows:  In section II we give a brief review of a general formalism to describe free spinning particles with arbitrary spin. In section III we focus on the spin-1 particle: introducing background Yang-Mills interaction to the theory and deriving the vertex operator for the external Yang-Mills fields. In section IV we define the vacuum, ghost measure and Green functions for Yang-Mills tree amplitudes. In section V we present the calculation of 3 and 4-point trees, and one-loop amplitudes, using the worldline approach, since it is sufficient for these amplitudes. In section VI we discuss the worldgraph approach that follows string calculations more closely, and show how it can reproduce the tree results derived from the worldline approach.

\section{II. Free spinning particles}

We begin with the free BRST charge for arbitrary spin. A useful method for deriving gauge invariant actions is the OSp(1,1$|$2) formalism \cite{Siegel:1986zi}, where one starts with the light-cone SO(D$-$2) linearly realized by the physical states, and adds two bosonic coordinates to restore Lorentz covariance and two fermionic coordinates to cancel the additional degrees of freedom. Thus the $\rm{SO}(D-2)$ representation is extended to $\rm{OSp}(D-1,1|2)$, and the non-linearly realized $\rm{SO}(D-1,1)$ of the physical states is extended to $\rm{OSp}(D,2|2)$. The action then uses only the subgroup $\rm{SO}(D-1,1)\otimes \rm{OSp}(1,1|2)$, where the $\rm{OSp}(1,1|2)$ is a symmetry of the unphysical (orthogonal) directions under which the physical states should be singlets (in the cohomology). We use $(A,B...)$ for $\rm{OSp}(D,2|2)$ indices, $(a,b...)$ for the $\rm{SO}(D-1,1)$ part and $(+,-)$, $(\oplus,\ominus)$ for the bosonic and fermionic indices of $\rm{OSp}(1,1|2)$ respectively. The easiest way is then to begin with linear generators $J^{AB}$ of $\rm{OSp}(D,2|2)$, use the gauge symmetry to gauge away the + direction of $\rm{OSp}(1,1|2)$ and use equations of motion to fix the $-$ direction. Then the kinetic operator of the action is simply the delta function of the $\rm{OSp}(1,1|2)$ generators (now non-trivial due to solving the equation of motion).

One can further simplify things by utilizing only a subset of the generators of $\rm{OSp}(1,1|2)$.  (This is analogous to the method of finding $\rm{SU}(2)$ singlets by looking at states annihilated by $J_{3}$ and $J_{-}$.)  In the end one is left with the group $\rm{IGL}(1)$ with generators $J^{\oplus\ominus}$ and $J^{\oplus -}$. Relabeling $c=x^{\oplus}$ and $b=\partial_{\oplus}$,
\begin{equation}\label{1}
J=iJ^{\oplus\ominus}+1=cb+iS^{\oplus\ominus},\quad Q=J^{\oplus -}={\textstyle{1 \over 2}}c\partial ^2 + S^{ \oplus a} \partial _a  + S^{ \oplus  \oplus } b
\end{equation}
$J$ will be the ghost number and $Q$ the BRST charge. One is then left with the task of finding different representation for $S^{AB}$ satisfying the algebra
\[
\left[S_{AB},S^{CD}\right\}=-\delta_{[A}^{[C} S_{B\}}{}^{D\}}
\]
There may be more than one representation corresponding to the same spin. It is easy to build massless spin-$\frac{1}{2}$ representations using gamma matrices
\[
\text{spin-} \textstyle{1 \over 2}: \quad\quad S_{AB}=-\textstyle{1 \over 2}[\gamma_{A}.\gamma_{B}\},\quad\{\gamma_{A},\gamma_{B}]=-\eta_{AB}
\]
and spin-1 using ket-bra
\[
\text{spin-}1:\quad\quad S_{AB}=|_{[A}\rangle \langle _{B\}}|,\quad \left\langle_{A}|_{B}\right\rangle=\eta_{AB}
\]
All higher spins can be built out of these two. For a review of the $\rm{OSp}(1,1|2)$ formalism see \cite{Siegel:1999ew}.

For our purpose we use first-quantized fields (i.e. fields on a worldline) to form representations. It is known that the free relativistic spin-$\frac{N}{2}$ particle can be described by a first-quantized action with $N$-extended worldline supersymmetry \cite{Brink:1976uf}. For example, for spin $\frac{1}{2}$ we use $N=1$ worldline fields $\psi ^A$ where $\psi^a$ are fermionic fields and $\psi^\oplus=i\gamma,\,\psi^\ominus=i\beta$ are the bosonic ghosts for SUSY. We summarize this representation as follows
\begin{eqnarray}
 S^{ab}  &=& {\textstyle{1 \over 2}}\left[ {\psi ^a ,\psi ^b } \right] = \psi ^a \psi ^b  \\
\nonumber S^{ \oplus a}  &=& {\textstyle{i \over 2}}\left\{ {\gamma ,\psi ^a } \right\} = i\gamma \psi ^a  \\
\nonumber S^{ \oplus  \oplus }  &=& {\textstyle{1 \over 2}}\left\{ {\gamma ,\gamma } \right\} = \gamma ^2
\end{eqnarray}
and
\[
\begin{array}{c}
 \{\psi ^a ,\psi ^b\} = \eta ^{ab}  \\
 \left[ {\gamma ,\psi ^a } \right] = 0 \\
 \left[ {\gamma ,\gamma } \right] = 0 \\
 \end{array}
\]
In this letter we focus on the $N=2$ spinning particle representation for massless vector states. Now, due to $N=2$ there are a pair of worldline spinors $\left(\psi^{a},\bar{\psi}^{b}\right)$ and similarly bosonic ghosts $\left(\gamma, \bar{\gamma}, \beta, \bar{\beta}\right)$.
The spin operators are then:
\begin{eqnarray}
 S^{ab}  &=& \bar \psi ^a \psi ^b  - \bar \psi ^b \psi ^a  \\
\nonumber S^{ \oplus a}  &=& i\gamma \bar \psi ^a  + i\bar \gamma \psi ^a  \\
\nonumber S^{ \oplus  \oplus }  &=& 2\gamma \bar \gamma
 \end{eqnarray}
with the following (anti-)commutation relations for the fields:
\[
\{ \bar{\psi}^{a} ,\psi^{b}\} = \eta ^{ab}
\]
\[
\{\bar{\psi}^a ,\bar{\psi}^b\} = \{\psi^a ,\psi^b\} = [\gamma ,\beta] =[\bar{\gamma} ,\bar{\beta }] =0
\]
\[
[ \gamma ,\bar{\beta}] = [\bar{\gamma} ,\beta] =\{ b,c\} = 1
\]

\section{III. Interacting spinning particles}

Interaction with external fields is introduced by covariantizing all the derivatives in the free BRST charge and adding a term proportional to $iF_{ab} S^{ab}$, which is the only term allowed by dimension analysis and Lorentz symmetry. The relative coefficient can be fixed by requiring the new interacting BRST charge $Q_{\rm{I}}$ to be nilpotent. In general the result is:
\begin{equation}\label{2}
Q_{\rm{I}}  = {\textstyle{1 \over 2}}c\left( {\nabla ^2  + iF_{ab} S^{ab} } \right) + S^{ \oplus a} \nabla _a  + S^{ \oplus  \oplus } b
\end{equation}
where we use the following convention for the covariant derivative and the field strength:
\[
\nabla _a  \equiv \partial _a  + iA_a
\]
\[
iF_{ab} \equiv \left[ {\nabla _a ,\nabla _b } \right]
\]

The nilpotency of $Q_{\rm I}$ can be used to derive vertex operators that are $Q$ closed. If we define the vertex operator as
\[
V = Q_{\rm I}  - Q
\]
Then
\[
Q_{\rm I}^{\rm{2}}  = 0 \Rightarrow \left\{ {Q,V} \right\} + V^2  = 0
\]
In the linearized limit, which is relevant for asymptotic states, we take only the part of $V$ that is linear in background fields (denoted by $V_{0}$). Then one has
\[
\left\{ {Q,V_{0}} \right\} = 0
\]

There will be an additional U(1) symmetry in the $N=2$ model. The vector states should be U(1) singlets and can be picked out by multiplying the original $Q_{\rm{I}}$ in eq.\ (\ref{2}) with an additional $\delta$ function (a U(1) projector).
\[Q_{\rm{I}}'=\delta\left(J_{\rm{U(1)}}\right)Q_{\rm{I}}\]
$J_{\rm{U(1)}}$ is the U(1) current:
\[
J_{\rm{U(1)}} = {\textstyle{1 \over 2}}\left( {\psi  \cdot \bar \psi  - \bar \psi  \cdot \psi } \right) - \gamma \bar \beta  + \bar \gamma \beta  =  - \bar \psi  \cdot \psi  + {\textstyle{D \over 2}} - \gamma \bar \beta  + \bar \gamma \beta
\]
where $D$ is the spacetime dimension and $\bar \psi ^a$,  $\bar \gamma$,  $\bar \beta$ have $\rm{U(1)}$  charge $-1$,  and their complex conjugates have $+1$. This $\rm{U(1)}$ constraint is important in that it ensures that $Q_{\rm{I}}$ for the $N=2$ model is indeed nilpotent. We will show this is the case.

Before choosing any specific representation, we have
\begin{eqnarray}
Q_{\rm{I}}^{\prime 2}&=&\delta\left(J_{\rm{U(1)}}\right)\,Q_{\rm{I}}^{2}=\delta\left(J_{\rm{U(1)}}\right)\,\textstyle{1 \over 2} \left\{Q_{\rm{I}} ,Q_{\rm{I}}\right\} \\
\nonumber&=&\delta\left(J_{\rm{U(1)}}\right)\textstyle{1 \over 2}\left\{ -icS^{ \oplus a}[ {\nabla ^b ,F_{ab} }] - icS^{ \oplus c} S^{ab} \left[ {\nabla _c ,F_{ab} } \right] + iS^{ \oplus  \oplus } S^{ab} F_{ab}  + iS^{ \oplus a} S^{ \oplus b} F_{ab} \right\}
 \end{eqnarray}

To understand how the projector works for the $N=2$ model, consider normal ordering with respect to the following scalar vacuum:
\[
\left(\gamma,\,\beta,\,\psi,\, b\right)\left| {\rm{0}} \right\rangle  =0
\]
This vacuum has $\rm{U(1)}$ charge $+1$ . A general normal-ordered operator with  $\ge 2$ barred fields on the left (unbarred fields are on the right), acting on any state built from the above vacuum, will either vanish or have negative $\rm{U(1)}$  charge. Therefore normal-ordered operators with $\ge 2$  barred fields will be projected out by $\delta (J_{\rm{U(1)}})$. Actually this property can be made true for any vacuum:  One just needs to shift the current by a constant in the projection operator.

With this in mind we have the following:
\begin{eqnarray}
\delta \left( J_{\rm{U(1)}} \right)S^{ \oplus a} S^{ \oplus b}  &=& \delta \left( J_{\rm{U(1)}} \right)\left( {i\gamma \bar \psi ^a  + i\bar \gamma \psi ^a } \right)\left( {i\gamma \bar \psi ^b  + i\bar \gamma \psi ^b } \right) =  - \delta \left( J_{\rm{U(1)}}\right)\bar \gamma \gamma \eta ^{ab}=-\delta \left( J_{\rm{U(1)}}\right)\frac{1}{2}S^{\oplus\oplus}\eta ^{ab}\nonumber\\
\delta \left( J_{\rm{U(1)}} \right)S^{ \oplus  \oplus } S^{ab}  &=& \delta \left( J_{\rm{U(1)}} \right)2\gamma \bar \gamma \left( {\bar \psi ^a \psi ^b  - \bar \psi ^b \psi ^a } \right) = 0\\
\nonumber\delta \left( J_{\rm{U(1)}} \right)S^{ \oplus c} S^{ab}  &=& \delta \left( J_{\rm{U(1)}} \right)\left( {i\gamma \bar \psi ^c  + i\bar \gamma \psi ^c } \right)\left( {\bar \psi ^a \psi ^b  - \bar \psi ^b \psi ^a } \right) =  \delta \left( J_{\rm{U(1)}}\right)\left({i\bar \gamma \psi ^b \eta ^{ac}  - i\bar \gamma \psi ^a \eta ^{bc} }\right)
\end{eqnarray}
Note that one could arrive at the same algebra for the spin operators if one were to use the spin-1 ket-bra representation introduced in the previous section; thus one again sees that the U(1) projector acts as picking out vector states. In fact the nilpotency of the BRST charge can be checked more easily using the ket-bra representation; however, for completeness we plug the above result into our previous calculation for $Q_{I}^{2\prime}$.  We have
\[
\begin{array}{c}
 \delta \left( J_{\rm{U(1)}} \right)Q_{\rm{I}}^2  =  c\delta \left( J_{\rm{U(1)}}\right)\left( {\bar \psi ^a \gamma  - \bar \gamma \psi ^a } \right)\left[ {\nabla ^b ,F_{ab} } \right] \\
 \end{array}
\]
which is proportional to the equation of motion satisfied by the asymptotic states. So we have proved that $\delta \left( J_{\rm{U(1)}} \right)Q_{\rm{I}}^2  = 0$.

The vertex operator is then easily obtained by considering $Q_{\rm I}$ as an expansion of $Q$,
\begin{eqnarray}\label{3}
 V_0 &=& \left[ Q_{\rm I}  - Q \right]_{\text{linear in $A$}} \\
\nonumber  &\equiv& cW_{\rm{I}}  + W_{{\rm{II}}}  \\
\nonumber  &=& {\textstyle{1 \over 2}}c\left[ {2 i A \cdot \partial + i\left(\partial_a A_b-\partial_b A_a\right) S^{ab} } \right] + i A_a S^{ \oplus a} \\
\nonumber  &=& - \epsilon _a \left[ {c\left( {i\dot X^a  + \bar \psi ^b \psi ^a k_b  - \bar \psi ^a \psi ^b k_b } \right) + \left( {\gamma \bar \psi ^a  + \bar \gamma \psi ^a } \right)} \right]\exp \left[ {ik \cdot X\left( \tau  \right)} \right]
\end{eqnarray}
This vertex operator satisfies
\[
\left\{ {Q,V_0} \right\} = 0
\]
The integrated vertex can be derived by noting:
\[[Q,W_{\rm{I}}]=\partial V_0\,\rightarrow\,\left[Q,\int W_{\rm{I}}\right]=0\]

More complicated vertex operators are needed for the usual worldline formalism. We will discuss in detail how these operators arise in section V. In the worldgraph formalism linearized vertex operators derived above will be sufficient.

\section{IV. Vacuum, Ghost measure and Green Functions}

When calculating amplitudes, the vacuum with which one chooses to work dictates the form of vertex operator and insertions one needs. In string theory, different choices of vacuum are called different pictures. The scalar vacuum discussed above is defined by the expectation value
\[\left\langle0| c|0\right\rangle\sim1\]
The conformal vacuum of string theory
\[\left\langle0| ccc|0\right\rangle\sim1\]
does not exist in particle theory since there aren't that many zero modes to saturate at tree level. On the other hand one could also treat the worldline SUSY ghosts' zero modes, which would require additional insertions. These are defined by the vacuum
\[(\bar{\beta},\,\beta,\,\psi,\, b)\left| {\rm{\hat{0}}} \right\rangle =0\Rightarrow\left\langle\rm{\hat{0}}| c\delta(\gamma)\delta(\bar{\gamma})|\rm{\hat{0}}\right\rangle\sim1\]
which has U(1) charge 2 and is thus not a physical vacuum.

To use the vertex operator we found above, we need to find a $\rm{U}(1)$ neutral vacuum  $\left| {\tilde 0} \right\rangle$ that is in the cohomology of the free BRST charge $Q$. It is related to the previous vacuum through the following relation:
\[
\left| {\tilde 0} \right\rangle  = \bar \beta \left| 0 \right\rangle=\delta\left(\half\gamma^2\right)\left| \rm{\hat{0}} \right\rangle,
\]
which leads to
\[
\left\langle {\tilde 0} \right|\gamma c\bar \gamma \left| {\tilde 0} \right\rangle  \sim 1
\]
This vacuum can be understood as the Yang-Mills ghost. It has ghost number $-1$ and lies in the cohomology only at zero momentum, indicating a constant field. Therefore it corresponds to the global part of the gauge symmetry:   Gauge parameters satisfying $Q\Lambda=0$ have no effect on the gauge transformations in the free theory, $\delta\phi=Q\Lambda$. In principal one could proceed to compute amplitudes in the available vacua mentioned above; however, due to its U(1) neutral property, the Yang-Mills ghost vacuum should be the easiest to extend to higher loops, since it would be easier to enforce U(1) neutrality.

With the above definition of the vacuum and the ghost measure, we can easily obtain the tree-level Green function. For the worldline formalism the Green function for the $X$ fields at tree level is as usual,
\[
\eta ^{ab} G_{\rm{B}} \left( {\tau ,\tau '} \right) \equiv \left\langle {X^a \left( \tau  \right)X^b \left( {\tau '} \right)} \right\rangle  =  - \frac{1}{2}\eta ^{ab} \left| {\tau  - \tau '} \right|
\]
For the fermions:
\[
\eta^{ab} G_F(\tau , \tau')  \equiv \left\langle {\psi ^a \left( \tau  \right)\bar \psi ^b \left( {\tau '} \right)} \right\rangle  = \eta ^{ab} \Theta \left( {\tau  - \tau '} \right)
\]
where $\Theta$ is a step function which is zero if the argument is negative. Note that the fermionic Green function does not have the naive relation with the bosonic Green function
\[G_F \neq -\dot{G}_{\rm{B}}=\frac{1}{2} \text{sign} (\tau  - \tau ')\]
It differs by a constant $\frac{1}{2}$. This is due to different boundary conditions:  The vacuum we choose, which is at $t=-\infty$, is defined to be annihilated by $\psi^{a}$; therefore on a time ordered line the expectation can be non-vanishing only if $\psi$ is at later time then $\bar{\psi}$.

\section{V. Scattering Amplitudes (worldline approach)}

In the worldline approach, one starts by choosing a specific worldline, and then inserts relevant vertex operators for external states. For YM theory, where the worldline state is the same as external states, namely a vector, the choice for worldline is less obvious. Previous work on the worldline formalism was geared toward the calculation of one-loop amplitudes, where the loop itself provides a natural candidate for the worldline. This advantage is not present for tree or higher-loop amplitudes. Furthermore, for higher-point tree graphs, calculating the amplitude from the worldline requires sewing lower-point tree amplitudes to the worldline. This is unsatisfactory from the viewpoint of first-quantized perturbation theory.   

In general, to calculate an $n$-point tree-level partial amplitude in the worldline approach:

1. Choose a specific color ordering (e.g., $1$$2$...$n$). Label external lines counter-clockwise.

\begin{figure}
\includegraphics{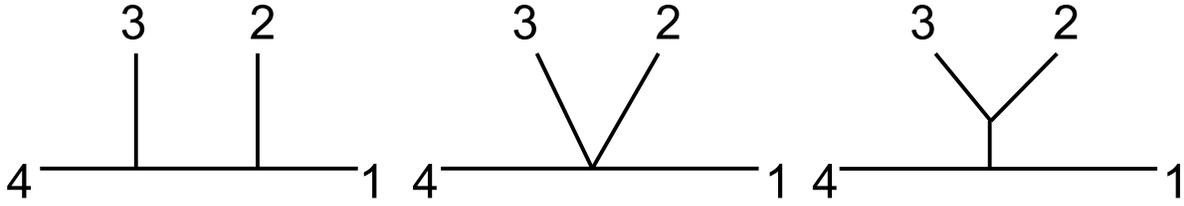}
\caption
{
    \label{FourPointsWL1}
    Three diagrams to be calculated if one chooses to connect lines 1 and 4 as the worldline. The second diagram needs a pinch operator, and the third diagram needs a vertex operator representing the tree attached to the worldline.
}
\end{figure}

2. Draw a worldline between any two of the external lines (e.g., line $1$ and line $n$) and connect all other external lines to this worldline in the following three ways: (a) Use the linearized vertex operator $V_{0}$ defined in section $\rm{III}$. (b) Use a vertex operator that is quadratic in background fields (``pinching"). This quadratic vertex operator (``pinch operator") can be derived from eq.(\ref{3}) by extending the field strength to contain the non-abelian terms and takes the form  
\[
v^{(ij)} = \epsilon_{ia} \epsilon_{jb} c\left( {\bar \psi ^b \psi ^a  - \bar \psi ^a \psi ^b } \right)e^{i\left(k_{i}+k_{j}\right)\cdot X}
\]
(c) Have the external lines first form a lower-point tree graph and then connect to the worldline through either of the two vertex operators mentioned previously. This corresponds to replacing $A^{a}=\epsilon^{a}e^{ik\cdot X}$ with the non-linear part of the solution to the field equations that the background field satisfies. For example, for a four-point tree amplitude there are the three graphs shown in fig.\ (\ref{FourPointsWL1}), representing the three different ways external fields can attach to the worldline. 

For lower-point graphs it is possible to choose the worldline in such a way that only linear vertex operators are required. We will show this in our actual computation for the four-point amplitude.

3. For each of the diagrams from above, insert three fixed vertex operators (respectively fixed at $\tau=\infty, 0, -\infty$). Two of them represent the initial and final external states that were connected to form the worldline, while the remaining one can be any of the operators described above. For example, one has:
\[
V^{(n)}(\infty)V^{(2)}(0)V^{(1)}(-\infty) \text{ or } V^{(n)}(\infty)v^{(32)}(0)V^{(1)}(-\infty)
\]
where the superscript $(i)$ represents the momentum and polarization vector of the external line $i$.

4. Insert the remaining vertex operators as the integrated ones, e.g.,
\[
\int{W^{(i)}_{{\rm I}}} \text{ or } \int{v^{(ij)}}
\]
with the integration regions so chosen that the diagram is kept planar.

5. Evaluate the expectation value with respect to the Yang-Mills ghost vacuum. 

The fact that one needs to calculate lower-point tree graphs for a general tree graph is unsatisfactory, since one should be able to calculate an arbitrary-point amplitude without the knowledge of its lower-point counterparts. This was less of a problem in the previous one-loop calculations, since one can claim that the method was really for 1PI graphs, and therefore sewing is necessary to calculate graphs that are not 1PI. It is more desirable to be able to calculate any amplitude with the knowledge of just the vertex operators and Green functions. This will be the aim of the ``worldgraph'' approach, which we leave to section VI. We first proceed to show how to calculate 3- and 4-point trees, and one-loop amplitudes, by the worldline approach. 

\subsection{1. 3-Point Tree}

In the 3-point case, we connect line 1 and line 3 as the worldline. The three vertex operators are respectively fixed at $\tau _{C} \to \infty$, $\tau _{B}  = 0$ and $\tau _{A}  \to  - \infty$. Note that we need one $c$ ghost to saturate the zero-mode and give a non-vanishing expectation value:
\begin{eqnarray}
 A_3  &=& \left\langle {V^{(3)}\left( \tau_{C} \right)V^{(2)}\left( {\tau_{B} } \right)V^{(1)}\left( \tau_{A} \right)} \right\rangle  \\
  \nonumber &=& \left\langle {\left[ {cW^{(3)}_{{\rm I}} \left( {\tau_{C}} \right)} \right]\left[ {W^{(2)}_{{\rm II}} \left( {\tau_{B}} \right)} \right]\left[ {W^{(1)}_{{\rm II}} \left( {\tau_{A}} \right)} \right]} \right\rangle  \\
  \nonumber &+& \left\langle {\left[ {W^{(3)}_{{\rm II}} \left( {\tau_{C}} \right)} \right]\left[ {cW^{(2)}_{{\rm I}} \left( {\tau_{B}} \right)} \right]\left[ {W^{(1)}_{{\rm II}} \left( {\tau_{A}} \right)} \right]} \right\rangle  \\
  \nonumber &+& \left\langle {\left[ {W^{(3)}_{{\rm II}} \left( {\tau_{C}} \right)} \right]\left[ {W^{(2)}_{{\rm II}} \left( {\tau_{B}} \right)} \right]\left[ {cW^{(1)}_{{\rm I}} \left( {\tau_{A}} \right)} \right]} \right\rangle
 \end{eqnarray}
The first term and the third term vanish due to $\epsilon\cdot\dot X$ in $W_{\rm{I}}$ contracting with the $e^{ik\cdot X}$ in the other two $W_{\rm{II}}$'s, which are proportional to $\epsilon _3  \cdot k_3$ and $\epsilon _1  \cdot k_1$ respectively, and vanish in the Lorenz gauge. The remaining term becomes
\begin{eqnarray}
 A_3 &=& \left\langle {\left[ {W^{(3)}_{{\rm II}} \left( {\tau_{C}} \right)} \right]\left[ {cW^{(2)}_{{\rm I}} \left( {\tau_{B}} \right)} \right]\left[ {W^{(1)}_{{\rm II}} \left( {\tau_{A}} \right)} \right]} \right\rangle  \\
\nonumber     &=& - \epsilon_{3a}\epsilon_{2c}\epsilon_{1d}\left\langle \left[\left(\gamma\bar{\psi}^{a}+\bar{\gamma}\psi^{a}\right)e^{ik_3 \cdot X}\right]_{\tau_{C}}
c\left[\left( k_1^c+(\bar{\psi}^{b}\psi^{c}-\bar{\psi}^{c}\psi^{b})k_{2b}\right)e^{ik_2 \cdot X}\right]_{\tau_{B}}
\left[\left(\gamma\bar{\psi}^{d}+\bar{\gamma}\psi^{d}\right)e^{ik_1 \cdot X}\right]_{\tau_{A}}\right\rangle\\
\nonumber     &=& - \left[ (\epsilon_3\cdot\epsilon_1)(\epsilon_2\cdot k_3) + (\epsilon_1\cdot\epsilon_2)(\epsilon_3\cdot k_1)  + (\epsilon_2\cdot\epsilon_3)(\epsilon_1\cdot k_2) \right]
\end{eqnarray}
As usual (see, e.g., \cite{Dai:2006vj}), the contractions among the exponentials give an overall factor of $e^{ { - \sum_{A \le i < j \le C} {k_i  \cdot k_j G_B \left( {\tau _i  - \tau _j } \right)} } }$ in the final result, but this factor equals $1$ if we go on-shell.

\subsection{2. 4-Point Tree}

For the 4-point amplitude (with color-ordering $1234$), one can calculate the three diagrams in fig.\ (\ref{FourPointsWL1}), but as we have mentioned, one can simplify the calculation by choosing a worldline between line $1$ and line $3$. In this case, there is only one diagram to be calculated (fig.\ (\ref{FourPointsWL2})), and there is only one integrated vertex operator --- line 4. We fix the other three as $\tau_{\rm{D}}  \to \infty$, $\tau _{\rm{C}}=0$ and $\tau _{\rm{A}}  \to  - \infty$, and the integrated vertex has integration region $\tau_{\rm{D}}\geq\tau _{\rm{B}}\geq\tau _{\rm{A}}$. We then have:
\begin{figure}
\includegraphics{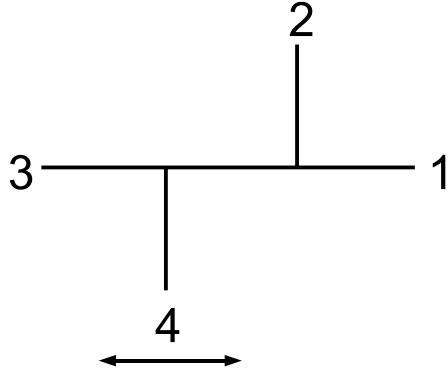}
\caption
{
    \label{FourPointsWL2}
    If one chooses to connect line 1 and line 3 as the worldline, there is only one diagram to be calculated. There is no need for pinch or more complicated operators. Note that line 4 is the integrated vertex and the integration region can be from $-\infty$ to $+\infty$, still keeping the graph planar.
}
\end{figure}
\begin{eqnarray}
 A_4  &=& \left\langle {\left[ {V^{(3)} \left( {\tau_{D}} \right)} \right]\left[ {V^{(4)} \left( {\tau_{C}} \right)} \right]\left[ {\int_{\tau_{A}}^{\tau_{D}} {W_{{\rm{I}}}^{{\rm{(2)}}} \left( {\tau_{B}} \right)d\tau _B } } \right]\left[ {V^{(1)} \left( {\tau_{A}} \right)} \right]} \right\rangle  \\
\nonumber  &=& \left\langle {\left[ {cW_{\rm{I}}^{{\rm{(3)}}} \left( {\tau_{D}} \right)} \right]\left[ {W_{{\rm{II}}}^{{\rm{(4)}}} \left( {\tau_{C}} \right)} \right]\left[ {\int_{\tau_{A}}^{\tau_{D}} {W_{{\rm{I}}}^{{\rm{(2)}}} \left( {\tau_{B}} \right)d\tau _B } } \right]\left[ {W_{{\rm{II}}}^{{\rm{(1)}}} \left( {\tau_{A}} \right)} \right]} \right\rangle  \\
\nonumber  && +\left\langle {\left[ {W_{{\rm{II}}}^{{\rm{(3)}}} \left( {\tau_{D}} \right)} \right]\left[ {cW_{\rm{I}}^{{\rm{(4)}}} \left( {\tau_{C}} \right)} \right]\left[ {\int_{\tau_{A}}^{\tau_{D}} {W_{{\rm{I}}}^{{\rm{(2)}}} \left( {\tau_{B}} \right)d\tau _B } } \right]\left[ {W_{{\rm{II}}}^{{\rm{(1)}}} \left( {\tau_{A}} \right)} \right]} \right\rangle  \\
\nonumber  && +\left\langle {\left[ {W_{{\rm{II}}}^{{\rm{(3)}}} \left( {\tau_{D}} \right)} \right]\left[ {W_{{\rm{II}}}^{{\rm{(4)}}} \left( {\tau_{C}} \right)} \right]\left[ {\int_{\tau_{A}}^{\tau_{D}} {W_{{\rm{I}}}^{{\rm{(2)}}} \left( {\tau_{B}} \right)d\tau _B } } \right]\left[ {cW_{\rm{I}}^{{\rm{(1)}}} \left( {\tau_{A}} \right)} \right]} \right\rangle
 \end{eqnarray}
The first and third term again vanish, for the same reason as in the three-point case. The remaining term can be written in two parts by separating the integration region:
\begin{eqnarray}
 A_4  &=& A_{4s}  + A_{4t}  \\
  \nonumber &=& \left\langle {\left[ {W_{{\rm{II}}}^{{\rm{(3)}}} \left( {\tau_{D}} \right)} \right]\left[ {cW_{\rm{I}}^{{\rm{(4)}}} \left( {\tau_{C}} \right)} \right]\left[ {\int_{\tau_{A}}^{\tau_{C}} {W_{{\rm{I}}}^{{\rm{(2)}}} \left( {\tau_{B}} \right)d{\tau_{B}}} } \right]\left[ {W_{{\rm{II}}}^{{\rm{(1)}}} \left( {\tau_{A}} \right)} \right]} \right\rangle  \\
 \nonumber && +\left\langle {\left[ {W_{{\rm{II}}}^{{\rm{(3)}}} \left( {\tau_{D}} \right)} \right]\left[ {\int_{\tau_{C}}^{\tau_{D}} {W_{{\rm{I}}}^{{\rm{(2)}}} \left( {\tau_{B}} \right)d{\tau_{B}}} } \right]\left[ {cW_{\rm{I}}^{{\rm{(4)}}} \left( {\tau_{C}} \right)} \right]\left[ {W_{{\rm{II}}}^{{\rm{(1)}}} \left( {\tau_{A}} \right)} \right]} \right\rangle
 \end{eqnarray}
Actually one can see these two terms as representing the $s$-channel and $t$-channel graphs from the second-quantized approach (see fig.\ (\ref{TwoRegions})). The $\tau$'s are time ordered according to the order they appear on the worldline. The results are:
\begin{figure}
\includegraphics{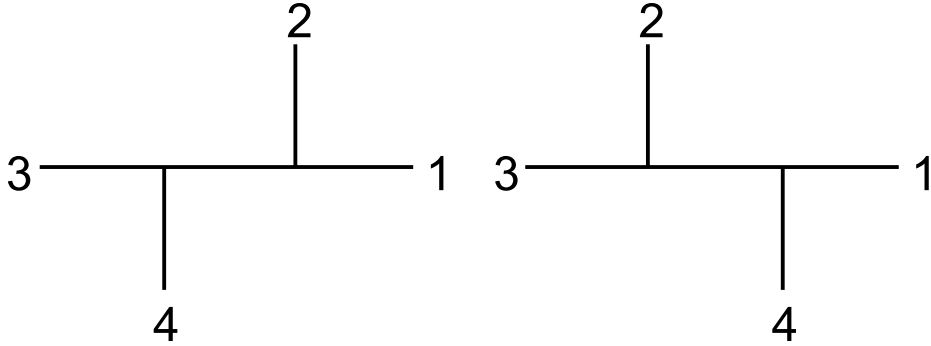}
\caption
{
    \label{TwoRegions}
    Two integration regions.  The integrated vertex sits at 2.
}
\end{figure}
\[
A_{4s}  =  - \frac{2}{s}\left[ \begin{array}{l}
  - \frac{s}{4}\left( {\epsilon _1  \cdot \epsilon _3 } \right)\left( {\epsilon _2  \cdot \epsilon _4 } \right) - \frac{u}{2}\left( {\epsilon _1  \cdot \epsilon _2 } \right)\left( {\epsilon _4  \cdot \epsilon _3 } \right) \\
  + \left( {\epsilon _2  \cdot k_1 } \right)\left( {\epsilon _4  \cdot k_3 } \right)\left( {\epsilon _1  \cdot \epsilon _3 } \right) + \left( {\epsilon _1  \cdot k_2 } \right)\left( {\epsilon _3  \cdot k_4 } \right)\left( {\epsilon _2  \cdot \epsilon _4 } \right) \\
  + \left( {\epsilon _1  \cdot k_3 } \right)\left( {\epsilon _2  \cdot k_4 } \right)\left( {\epsilon _3  \cdot \epsilon _4 } \right) + \left( {\epsilon _4  \cdot k_2 } \right)\left( {\epsilon _3  \cdot k_1 } \right)\left( {\epsilon _1  \cdot \epsilon _2 } \right) \\
  - \left( {\epsilon _1  \cdot k_2 } \right)\left( {\epsilon _4  \cdot k_3 } \right)\left( {\epsilon _2  \cdot \epsilon _3 } \right) - \left( {\epsilon _3  \cdot k_4 } \right)\left( {\epsilon _2  \cdot k_1 } \right)\left( {\epsilon _1  \cdot \epsilon _4 } \right) \\
  - \left( {\epsilon _1  \cdot k_4 } \right)\left( {\epsilon _2  \cdot k_3 } \right)\left( {\epsilon _3  \cdot \epsilon _4 } \right) - \left( {\epsilon _3  \cdot k_2 } \right)\left( {\epsilon _4  \cdot k_1 } \right)\left( {\epsilon _1  \cdot \epsilon _2 } \right) \\
 \end{array} \right]
\]
\[
A_{4t}  =  - \frac{2}{t}\left[ \begin{array}{l}
  - {\textstyle{t \over 4}}\left( {\epsilon _1  \cdot \epsilon _3 } \right)\left( {\epsilon _2  \cdot \epsilon _4 } \right) - {\textstyle{u \over 2}}\left( {\epsilon _1  \cdot \epsilon _4 } \right)\left( {\epsilon _2  \cdot \epsilon _3 } \right) \\
  + \left( {\epsilon _1  \cdot k_4 } \right)\left( {\epsilon _3  \cdot k_2 } \right)\left( {\epsilon _2  \cdot \epsilon _4 } \right) + \left( {\epsilon _2  \cdot k_3 } \right)\left( {\epsilon _4  \cdot k_1 } \right)\left( {\epsilon _1  \cdot \epsilon _3 } \right) \\
  + \left( {\epsilon _1  \cdot k_3 } \right)\left( {\epsilon _4  \cdot k_2 } \right)\left( {\epsilon _2  \cdot \epsilon _3 } \right) + \left( {\epsilon _2  \cdot k_4 } \right)\left( {\epsilon _3  \cdot k_1 } \right)\left( {\epsilon _1  \cdot \epsilon _4 } \right) \\
  - \left( {\epsilon _1  \cdot k_2 } \right)\left( {\epsilon _4  \cdot k_3 } \right)\left( {\epsilon _2  \cdot \epsilon _3 } \right) - \left( {\epsilon _2  \cdot k_1 } \right)\left( {\epsilon _3  \cdot k_4 } \right)\left( {\epsilon _1  \cdot \epsilon _4 } \right) \\
  - \left( {\epsilon _1  \cdot k_4 } \right)\left( {\epsilon _2  \cdot k_3 } \right)\left( {\epsilon _3  \cdot \epsilon _4 } \right) - \left( {\epsilon _3  \cdot k_2 } \right)\left( {\epsilon _4  \cdot k_1 } \right)\left( {\epsilon _1  \cdot \epsilon _2 } \right) \\
 \end{array} \right]
\]
The sum of the above two parts is exactly the 4-point Yang-Mills tree amplitude. Note that we don't need the pinch operator in this calculation. This is because there cannot be a pinch operator representing line 2 and line 4, since they are not adjacent in the color ordering.

\subsection{3. One-Loop Amplitude}

It is straightforward to generalize this method to the calculation of 1-loop 1PI diagrams. The new feature in this case is that one must ensure U(1) neutrality inside the loop. One can think of the diagram as connecting both ends of a tree diagram, and only sum over U(1) neutral states. The U(1) neutral states are written as:
\[
\left| {A,p} \right\rangle  = \left\{ \begin{array}{l}
 \left| {a,p} \right\rangle  = \gamma \bar \psi ^a \left| {\tilde 0} \right\rangle  \otimes \left| p \right\rangle  \\
 \left| {{\rm{ghost}},p} \right\rangle  = \left| {\tilde 0} \right\rangle  \otimes \left| p \right\rangle  \\
 \left| {{\rm{antighost}},p} \right\rangle  = \gamma \bar \gamma \left| {\tilde 0} \right\rangle  \otimes \left| p \right\rangle  \\
 \end{array} \right.
\]
where $p$ is the momentum of the state, and the last two states are the Faddeev-Popov ghosts for background gauge fixing. The general expression for the amplitude of $n$-point 1-loop 1PI diagrams is then
\begin{eqnarray}
A_n^{{\rm{1 - loop}}}  &=& \sum\limits_{A,p} {\int_0^\infty  {dT\left\langle {A,p} \right|V^{(n)} \left( {\tau _n } \right)\prod\limits_{i = 1}^{n-1} {\int_{\tau _{i - 1}  \le \tau _i  \le \tau _{i + 1} } {d\tau _i W_{\rm{I}}^{(i)} \left( {\tau _i } \right)} } \left| {A,p} \right\rangle } } \\
\nonumber &+& \text{diagrams with pinch operators}
\end{eqnarray}
where we define $\tau _0 = 0$ and fix $\tau _n  = T$. Note that at one-loop  we don't have the freedom to choose worldline (it should always be the loop), so one cannot avoid using the pinch operators.

Another approach is to insert a U(1) projector in the loop to pick out all the U(1) neutral states. That is, one inserts:
\[
\delta \left[ {J_{\rm U(1)} } \right] = \frac{1}{2\pi}\int_0^{2\pi } {d\theta \exp \left[ {i \frac{\theta}{T} \int_0^T {d\tau J_{\rm U(1)}} } \right]}  = \frac{1}{2\pi}\int_0^{2\pi } {d\theta \exp \left[ {i\frac{\theta}{T} \int_0^T {d\tau ( - \bar \psi  \cdot \psi  + {\textstyle{D \over 2}} - \gamma \bar \beta  + \bar \gamma \beta)} } \right]}
\]
Similar approaches have been taken in \cite{Strassler:1992zr} and \cite{Bastianelli:2005vk}. In \cite{Strassler:1992zr}, $i\theta$ is interpreted as a mass to be taken to infinity at the end, and together with GSO-like projection kills all U(1) non-neutral states. For us the U(1) projector naturally gets rid of all unwanted states. Furthermore the worldline ghosts were not taken into account in \cite{Strassler:1992zr}; therefore they need to include the effect of Faddeev-Popov ghosts by adding covariant scalars to the action. This is sufficient for one loop, since they couple in the same way, yet will no longer be true for higher loops. Here we've (and also \cite{Bastianelli:2005vk}) included all the worldline ghosts; thus the Faddeev-Popov ghosts are naturally included. In \cite{Bastianelli:2005vk} gauge fixing the U(1) gauge field on a loop leads to a modulus, which is equivalent to $\theta$ in our U(1) projector insertion. The two views are analogous.

The inclusion of a U(1) projector amounts to additional quadratic terms in the action which will modify the Green function and introduce an additional $\theta$-dependent term to the measure. Here we give a brief discussion of its effect. The kinetic operator for the SUSY partners and SUSY ghosts is now:
\[
\partial _\tau + i\frac{\theta}{T}
\]
The $\theta$ term can be absorbed by redefining the U(1) charged fields,
\[
\begin{array}{*{20}c}
   {\Psi ' = e^{i\theta \tau /T} \Psi } & {\bar \Psi ' = e^{ - i\theta \tau /T} \bar \Psi }  \\
\end{array}
\]
where $\Psi = \left( {\psi ^a ,\gamma ,\beta } \right)$. Then the integration over $\theta$ is really integrating over all possible boundary conditions since:
\[\Psi '(T)=e^{i\theta}\Psi '(0)\]
Without loss of generality, we choose the periodic boundary condition for the original fields $\Psi$.

The 1-loop vacuum bubble is then computed through mode expansion on a circle with periodic boundary condition:
\[
{\rm{Det}} \left(\partial _\tau + i\frac{\theta}{T}\right)^{D-2}=\left[ {2i\sin \left( {{\textstyle{{\theta} \over 2}}} \right)} \right]^{D - 2}
\]
where $D$ comes from the $\psi \bar \psi$ integration and $-2$ comes from SUSY ghosts.
The fermionic Green function will be modified to
\[
G_F \left( {\tau ,\tau '} \right) = \frac{{e^{ -\frac{i\theta \left( {\tau  - \tau '} \right)}{T}} }}{{2i\sin {\textstyle{\theta  \over 2}}}}\left[ {e^{i{\textstyle{\theta  \over 2}}} \Theta \left( {\tau  - \tau '} \right) + e^{ - i{\textstyle{\theta  \over 2}}} \Theta \left( {\tau ' - \tau } \right)} \right]
\]
which satisfies the periodic boundary condition and differential equation
\[
\left( {\partial _\tau   + i\frac{\theta}{T} } \right)G_F \left( {\tau ,\tau '} \right) = \delta \left( {\tau  - \tau '} \right)
\]
Also, at one loop there are two zero-modes, one modulus (the circumference of the loop) and one Killing vector. The proper insertions for the vacuum are:
\[\left\langle\tilde{0}| bc|\tilde{0}\right\rangle\sim1\]

In general, the $n$-point 1-loop 1PI amplitude can thus be written as
\begin{eqnarray}
 A_n^{{\rm{1 - loop}}} & = & g^n\int_0^\infty  {\frac{1}{{T^{D/2} }}dT\left\langle {\delta \left[ {J_{{\rm{U(1)}}} } \right]bV^{(n)} \left( {\tau _n } \right)\prod\limits_{i = 1}^{n - 1} {\int_{\tau _{i - 1}  \le \tau _i  \le \tau _{i + 1} } {d\tau _i W_{\rm{I}}^{(i)} \left( {\tau _i } \right)} } } \right\rangle }  \\
 \nonumber &+& \text{diagrams with pinch operators} \\
  \nonumber &=& g^n\frac{1}{2\pi}\int_0^\infty  {\frac{1}{{T^{D/2} }}dT\int_0^{2\pi } {d\theta \left[ {2i\sin \left( {{\textstyle{{\theta} \over 2}}} \right)} \right]^{D - 2} } \left\langle {W_{\rm{I}}^{(n)} \left( {\tau _n } \right)\prod\limits_{i = 1}^{n - 1} {\int_{\tau _{i - 1}  \le \tau _i  \le \tau _{i + 1} } {d\tau _i W_{\rm{I}}^{(i)} \left( {\tau _i } \right)} } } \right\rangle } \\
  \nonumber &+& \text{diagrams with pinch operators}
 \end{eqnarray}
We've added the coupling constant $g$, but omitted group theory factors, such as a trace and a factor $\textstyle{N_c}$ of the number of colors for the planar contribution.  The $XX$ contraction should be calculated by the 1-loop bosonic Green function:
\[
\left\langle {X^a \left( \tau  \right)X^b \left( {\tau '} \right)} \right\rangle  = \eta ^{ab} G_B \left( {\tau  - \tau '} \right) = \eta ^{ab} \left[ { - \frac{1}{2}\left| {\tau  - \tau '} \right|+\frac{{\left( {\tau  - \tau '} \right)^2 }}{{2T}}} \right]
\]

For example, the two-point contribution to the effective action is (including the usual $-$ sign for the action, $\half$ for permutations, and group theory factor for this case)
\begin{eqnarray}
 \Gamma_2^{{\rm{1 - loop}}}  &=& -\frac{g^2N_c}{4\pi}\int_0^\infty  {\frac{1}{{T^{D/2} }}dT\int_0^{2\pi } {d\theta \left[ {2i\sin \left( {{\textstyle{\theta  \over 2}}} \right)} \right]^{D-2} } \int_{0}^{T}d\tau\left\langle W_{\rm{I}}^{(2)} \left( {T } \right){ W_{\rm{I}}^{(1)} \left( {\tau} \right)} \right\rangle }  \\
  \nonumber &=&-g^2N_c  \int_0^\infty  {\frac{1}{{T^{D/2} }}dT\int_0^T {d\tau \left[ \begin{array}{l}
 \left( { \delta \left( {T - \tau } \right) - {\textstyle{1 \over T}}} \right)\left( {\epsilon _1  \cdot \epsilon _2 } \right) \\
  +\left( {{\textstyle{1 \over 2}} - {\textstyle{\tau  \over T}}} \right)^2 \left( {\epsilon _2  \cdot k_1 } \right)\left( {\epsilon _1  \cdot k_2 } \right) \\
  - \left( {\epsilon _2  \cdot k_1 } \right)\left( {\epsilon _1  \cdot k_2 } \right) + \left( {k_1  \cdot k_2 } \right)\left( {\epsilon _1  \cdot \epsilon _2 } \right) \\
 \end{array} \right]} }  {e^{ {{\textstyle{1 \over 2}}k_1  \cdot k_2 \left( {T - \tau  - {\textstyle{{\left( {T - \tau } \right)^2 } \over T}}} \right)}}} \\
 \nonumber & = & -g^2N_c\left(\frac{k_1^2}{2}\right)^{-\epsilon}\left(1-\frac{1}{12}\right)\textstyle{\Gamma\left(\epsilon\right)}\left[(\epsilon_1\cdot\epsilon_2)(k_1\cdot k_2)-(\epsilon_1\cdot k_2)(\epsilon_2\cdot k_1)\right]\\
 \nonumber &=& -\frac{11}{24} {\rm tr}\ \left\{F_1^{ab}\left[{1 \over \epsilon} -\log\left(\half k_1^{2}\right)\right]F_{2ab}\right\}
\end{eqnarray}
In the final line we have used dimensional regularization $D=4-2\epsilon$, and dropped the term with the $\delta$ function, which gives the tadpole contribution. Modified minimal subtraction was used, with the conventions of ref.\ \cite{Fields}. Note that the $-\frac{1}{12}$ piece comes from the scalar graph while the $1$ comes from terms with the fermion Green function. The diagram with pinch operator does not contribute in this case.

\section{VI. Worldgraph Approach}

As mentioned previously, it is desirable even for tree graphs to develop a formalism that does not require an identification of a worldline to which external states are attached. Intuitively such a formalism would require one to simply identify 1D topological spaces that connect the external lines. This idea is very similar to string theory calculations and goes back as far as 1974 \cite{Casalbuoni:1974pj}. The main challenge for this ``worldgraph approach" (following \cite{Holzler:2007xt}) is the definition of Green functions on these non-differentiable topological spaces (non-differentiable because at interacting points it is not locally $R^1$). Previously, for multi-loops such Green functions have been derived by a combination of one-loop Green functions and insertions:  See \cite{Schubert:2001he} for review. Recently in \cite{Dai:2006vj} a more straightforward way to derive multi-loop Green functions was developed for scalar particles using the electric circuit analog. (A similar approach was used in \cite{Holzler:2007xt}.) Since fermion Green functions are related to bosons through a derivative (up to additional terms due to choice of vacuum or boundary conditions), what remains is to consistently define derivatives on these 1D topological spaces. We will use tree graphs as our testing ground.

\begin{figure}
\includegraphics{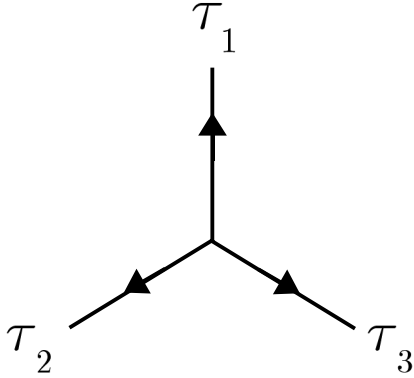}
\caption
{
    \label{ThreePoints}
   The topological space for a three-point interaction
}
\end{figure}

Consider the three-point amplitude:  One has only one graph, fig.\ (\ref{ThreePoints}). The arrows indicate the direction in which each $\tau_{i}$ is increasing. For scalar fields it was shown \cite{Dai:2006vj} that the appropriate Green function is proportional to the distance between two insertions; for the 3-point graph this is taken to be  $-\frac{1}{2}(\tau_{i}+\tau_{j})$. 

To define derivatives, one notes that they are worldline vectors and therefore must be conserved at each interaction point. This leads to the conclusion that if we denote the worldgraph derivative on each line as $D(\tau_{i})$, for the three-point graph they must satisfy:
\begin{equation}\label{4}
D_{\tau_{1}}+D_{\tau_{2}}+D_{\tau_{3}}=0
\end{equation}
This can be solved by defining the worldgraph derivatives as follows:
\begin{eqnarray}\label{5}
D_{\tau_1}=\partial_{\tau_2}-\partial_{\tau_3}\\
\nonumber D_{\tau_2}=\partial_{\tau_3}-\partial_{\tau_1}\\
\nonumber D_{\tau_3}=\partial_{\tau_1}-\partial_{\tau_2}
 \end{eqnarray}
There is another solution which corresponds to (counter-)clockwise orientation. The choice of orientation can be fixed by matching it with the color ordering. Since the derivative is a local operator, its definition will not be altered if the three-point graph is connected to other pieces to form larger graphs. The fermionic Green function then follows from the bosonic by taking $\psi$ as a worldline scalar and $\bar\psi$ as a worldline vector:
\[
G_F(\tau_i,\tau_j)\equiv\langle\bar{\psi}(\tau_i)\psi(\tau_j)\rangle=2\langle D_{\tau_i} X (\tau_i)X(\tau_j)\rangle
\]
Armed with these two Green functions we can show how the three-point amplitude works.

\subsection{1. 3-Point Tree}

For the three-point tree graph fig.\ (\ref{ThreePoints}) we start with:
\begin{eqnarray}\label{6}
 A_3  &=& \left\langle {V^{(3)}\left( \tau_{3} \right)V^{(2)}\left( {\tau_{2} } \right)V^{(1)}\left( \tau_{1} \right)} \right\rangle  \\
  \nonumber&=& \left\langle {\left[ {cW^{(3)}_{\rm{I}} \left( {\tau_{3}} \right)} \right]\left[ {W^{(2)}_{{\rm{II}}} \left( {\tau_{2}} \right)} \right]\left[ {W^{(1)}_{{\rm{II}}} \left( {\tau_{1}} \right)} \right]} \right\rangle  \\
  \nonumber&& +\left\langle {\left[ {W^{(3)}_{{\rm{II}}} \left( {\tau_{3}} \right)} \right]\left[ {cW^{(2)}_{\rm{I}} \left( {\tau_{2}} \right)} \right]\left[ {W^{(1)}_{{\rm{II}}} \left( {\tau_{1}} \right)} \right]} \right\rangle  \\
  \nonumber&& +\left\langle {\left[ {W^{(3)}_{{\rm{II}}} \left( {\tau_{3}} \right)} \right]\left[ {W^{(2)}_{{\rm{II}}} \left( {\tau_{2}} \right)} \right]\left[ {cW^{(1)}_{\rm{I}} \left( {\tau_{1}} \right)} \right]} \right\rangle
 \end{eqnarray}
Now the worldline derivatives in $W_{\rm{I}}$ are replaced by worldgraph derivatives defined in eq.\ (\ref{5}) and they give:
\[
\left\langle i\epsilon_1 \cdot D_{\tau_{1}} X(\tau_1)e^{i\left[\sum_{i=1}^{3}k_i\cdot X(\tau_{i})\right]}\right\rangle=-(\epsilon_1\cdot k_3)
\]
\[
\left\langle i\epsilon_2 \cdot D_{\tau_{2}} X(\tau_2)e^{i\left[\sum_{i=1}^{3}k_i\cdot X(\tau_{i})\right]}\right\rangle=-(\epsilon_2\cdot k_1)
\]
\[
\left\langle i\epsilon_3 \cdot D_{\tau_{3}} X(\tau_3)e^{i\left[\sum_{i=1}^{3}k_i\cdot X(\tau_{i})\right]}\right\rangle=-(\epsilon_3\cdot k_2)
\]
The fermionic Green functions are (with $F_{ij}\equiv\langle\bar{\psi}(\tau_i)\psi(\tau_j)\rangle$):
\begin{equation}\label{7}
\begin{array}{c @{, \quad} c @{, \quad} c}
    F_{12}=-1 & F_{23}=-1 & F_{31}=-1 \\
    F_{21}=+1 & F_{32}=+1 & F_{13}=+1
\end{array}
\end{equation}
Using the above one can compute eq.\ (\ref{6}). The first term becomes:
\begin{eqnarray}
A_{3-1}&=&\left\langle {\left[ {cW_{\rm{I}} \left( {\tau_{3}} \right)} \right]\left[ {W_{{\rm{II}}} \left( {\tau_{2}} \right)} \right]\left[ {W_{{\rm{II}}} \left( {\tau_{1}} \right)} \right]} \right\rangle\\
\nonumber&=&- \epsilon_{3a}\epsilon_{2c}\epsilon_{1d}\langle c[iDX^{a}+(\bar{\psi}^{b}\psi^{a}-\bar{\psi}^{a}\psi^{b})k_{3b}]_{\tau_{3}}[\gamma\bar{\psi}^{c}+\bar{\gamma}\psi^{c}]_{\tau_{2}}[\gamma\bar{\psi}^{d}+\bar{\gamma}\psi^{d}]_{\tau_{1}}e^{ik_{1}\cdot X_{\tau_{1}}}e^{ik_{2}\cdot X_{\tau_{2}}}e^{ik_{3}\cdot X_{\tau_{3}}}\rangle\\
\nonumber&=&-\epsilon_{2c}\epsilon_{1d}\langle [-(\epsilon_3\cdot k_2)+(\bar{\psi}^{b}\psi^{a}-\bar{\psi}^{a}\psi^{b})\epsilon_{3a}k_{3b}]_{\tau_{3}}[\bar{\psi}^{c}(\tau_{2})\psi^{d}(\tau_{1})+\psi^{c}(\tau_{2})\bar{\psi}^{d}(\tau_{1})]\rangle\\
\nonumber&=&2(\epsilon_3\cdot k_2)(\epsilon_2\cdot\epsilon_1)+2(\epsilon_2\cdot k_1)(\epsilon_3\cdot\epsilon_1)+2(\epsilon_1\cdot k_3)(\epsilon_2\cdot\epsilon_3)
\end{eqnarray}
A similar derivation gives the second and third terms:
\begin{eqnarray}
A_{3-2}&=&\left\langle {\left[ {W_{\rm{II}} \left( {\tau_{3}} \right)} \right]\left[ {cW_{{\rm{I}}} \left( {\tau_{2}} \right)} \right]\left[ {W_{{\rm{II}}} \left( {\tau_{1}} \right)} \right]} \right\rangle\\
\nonumber&=& 2(\epsilon_3\cdot k_2)(\epsilon_2\cdot\epsilon_1)+2(\epsilon_2\cdot k_1)(\epsilon_3\cdot\epsilon_1)+2(\epsilon_1\cdot k_3)(\epsilon_2\cdot\epsilon_3)\\
\nonumber A_{3-3}&=&\left\langle {\left[ {W_{{\rm{II}}} \left( {\tau_{3}} \right)} \right]\left[ {W_{{\rm{II}}} \left( {\tau_{2}} \right)} \right]\left[ {cW_{\rm{I}} \left( {\tau_{1}} \right)} \right]} \right\rangle\\
\nonumber&=& 2(\epsilon_3\cdot k_2)(\epsilon_2\cdot\epsilon_1)+2(\epsilon_2\cdot k_1)(\epsilon_3\cdot\epsilon_1)+2(\epsilon_1\cdot k_3)(\epsilon_2\cdot\epsilon_3)
\end{eqnarray}
Note that the three terms are the same, which respects the symmetry of the graph.

\subsection{2. 4-point Tree}

\begin{figure}
\includegraphics{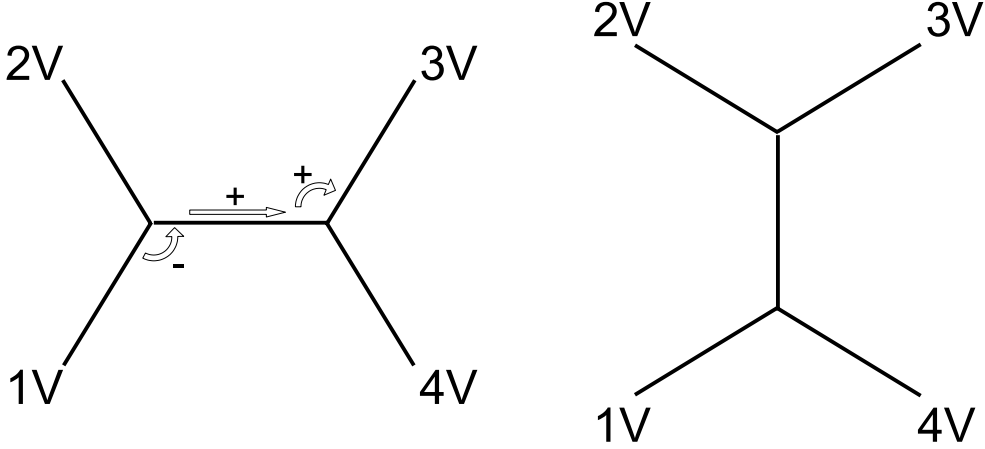}
\caption
{
    \label{FourPoints}
   The two graphs for the four-point interaction
}
\end{figure}

For the 4-point amplitude we have two graphs ($s$ channel and $t$ channel, see fig.\ (\ref{FourPoints})) constructed by connecting two three-point worldgraphs on a worldline. The worldline in the middle is actually a modulus of the theory, and one must insert a $b$ ghost. We focus on the $s$-channel graph; the $t$-channel graph can later be derived by exchanging the external momenta and polarizations in the $s$-channel amplitude. We wish to derive
\begin{eqnarray}\label{8}
A_{4s}&=&\int_{0}^{\infty}dT\left\langle  V^{(4)}(\tau_{4})V^{(3)}(\tau_{3})b(T)V^{(2)}(\tau_{2})V^{(1)}(\tau_{1})\right\rangle\\
\nonumber&=&\int_{0}^{\infty}dT
\left[ \begin{array}{r}
\left\langle  W_{{\rm II}}^{(4)}(\tau_{4})cW_{{\rm I}}^{(3)}(\tau_{3})b(T)cW_{{\rm I}}^{(2)}(\tau_{2})W_{{\rm II}}^{(1)}(\tau_{1})\right\rangle\\
+\left\langle  cW_{{\rm I}}^{(4)}(\tau_{4})cW_{{\rm I}}^{(3)}(\tau_{3})b(T)W_{{\rm II}}^{(2)}(\tau_{2})W_{{\rm II}}^{(1)}(\tau_{1})\right\rangle\\
+\left\langle  cW_{{\rm I}}^{(4)}(\tau_{4})W_{{\rm II}}^{(3)}(\tau_{3})b(T)cW_{{\rm I}}^{(2)}(\tau_{2})W_{{\rm II}}^{(1)}(\tau_{1})\right\rangle\\
+\left\langle  cW_{{\rm I}}^{(4)}(\tau_{4})W_{{\rm II}}^{(3)}(\tau_{3})b(T)W_{{\rm II}}^{(2)}(\tau_{2})cW_{{\rm I}}^{(1)}(\tau_{1})\right\rangle\\
+\left\langle  W_{{\rm II}}^{(4)}(\tau_{4})cW_{{\rm I}}^{(3)}(\tau_{3})b(T)W_{{\rm II}}^{(2)}(\tau_{2})cW_{{\rm I}}^{(1)}(\tau_{1})\right\rangle\\
+\left\langle  W_{{\rm II}}^{(4)}(\tau_{4})W_{{\rm II}}^{(3)}(\tau_{3})b(T)cW_{{\rm I}}^{(2)}(\tau_{2})cW_{{\rm I}}^{(1)}(\tau_{1})\right\rangle
\end{array}\right]
\end{eqnarray}

First we address the Green functions. As in \cite{Dai:2006vj} the bosonic Green function is still $-\frac{1}{2}L$, where $L$ is the length between two fields. Thus it is the same as in the three-point case, except that when the two fields sit on opposite ends of the modulus, one needs to add the value of the modulus $T$. The worldgraph derivatives still act the same way, since the definition is local, irrespective of other parts of the graph. This gives the following result for the $s$-channel graph:
\[\left\langle i\epsilon_1 \cdot D_{\tau_1}X(\tau_1)e^{i\left[\sum_{i=1}^{4}k_i\cdot X(\tau_{i})\right]}\right\rangle=-(\epsilon_1\cdot k_2)\]
\[\left\langle i\epsilon_2 \cdot D_{\tau_2}X(\tau_2)e^{i\left[\sum_{i=1}^{4}k_i\cdot X(\tau_{i}) \right]}\right\rangle=+(\epsilon_2\cdot k_1)\]
\[\left\langle i\epsilon_3 \cdot D_{\tau_3}X(\tau_3)e^{i\left[\sum_{i=1}^{4}k_i\cdot X(\tau_{i}) \right]}\right\rangle=-(\epsilon_3\cdot k_4)\]
\[\left\langle i\epsilon_4 \cdot D_{\tau_4}X(\tau_4)e^{i\left[\sum_{i=1}^{4}k_i\cdot X(\tau_{i}) \right]}\right\rangle=+(\epsilon_4\cdot k_3)\]

The fermionic Green functions are again more subtle. There are two types, that for $bc$ ghosts and that for the $\bar{\psi}\psi$. First one notes that on the modulus, which is a worldline, both Green functions should be a step function, as explained in section IV. This is sufficient for the $b,c$ ghosts. For $\bar{\psi}\psi$, since they can contract with each other on the same three-point graph or contract across the modulus, one must take the combined result:  For contractions on the same three-point graph the rules are just as eq.\ (\ref{7}), while for contraction across the modulus one multiplies the two Green functions on the two vertices with one from the modulus.  For example, in the $s$-channel graph fig.\ (\ref{FourPoints}):
\[\left\langle\bar{\psi}(\tau_1)\psi(\tau_3)\right\rangle=\left\langle\bar{\psi}(\tau_1)\psi(\tau_T)\right\rangle\left\langle\bar{\psi}(\tau_T)\psi(\tau_3)\right\rangle\Theta(T)=-1\]
As one can see, the contraction across the modulus is broken down as if there were a pair $\bar{\psi}\psi$ on each end of the modulus, contracting with the vertices separately, and a final step function due to the fact that the modulus is a worldline. (We choose the left time to be earlier.)  We now list all the relevant Green functions for the $s$-channel graph. The Green functions for the $bc$ ghosts are
\[
\begin{array}{c@{, \hfill\quad} c@{, \hfill\quad} c@{, \hfill\quad} c}
\left\langle c(\tau_1)b(T) \right\rangle=1 & \left\langle c(\tau_2)b(T) \right\rangle=1 & \left\langle c(\tau_3)b(T)\right\rangle=0 & \left\langle c(\tau_4)b(T) \right\rangle=0
\end{array}
\]
and the Green functions for the $\bar{\psi}\psi$ are (recall that we have defined $F_{ij}\equiv\langle\bar{\psi}(\tau_i)\psi(\tau_j)\rangle$)
\[
\begin{array}{c@{, \hfill\quad} c@{, \hfill\quad} c@{, \hfill\quad} c}
    F_{12}=+1 & F_{21}=-1 & F_{34}=+1 & F_{43}=-1 \\
    F_{23}=+1 & F_{32}=0  & F_{14}=+1 & F_{41}=0  \\
    F_{13}=-1 & F_{31}=0  & F_{24}=-1 & F_{42}=0
\end{array}
\]

Equipped with the Green functions one can compute eq.\ (\ref{8}). We do the $bc$ contractions first. Each term has two such contractions; using the above Green functions we see that the second and last terms cancel. We then have:
\[
A_{4s}=\int_{0}^{\infty} dT
\left[ 
\begin{array}{r}
 \left\langle cW_{\rm II}^{(4)}(\tau_{4})W_{\rm I}^{(3)}(\tau_{3})W_{\rm I}^{(2)}(\tau_{2})W_{\rm II}^{(1)}(\tau_{1})\right\rangle \\
-\left\langle cW_{\rm I}^{(4)}(\tau_{4})W_{\rm II}^{(3)}(\tau_{3})W_{\rm I}^{(2)}(\tau_{2})W_{\rm II}^{(1)}(\tau_{1})\right\rangle \\
+\left\langle cW_{\rm I}^{(4)}(\tau_{4})W_{\rm II}^{(3)}(\tau_{3})W_{\rm II}^{(2)}(\tau_{2})W_{\rm I}^{(1)}(\tau_{1})\right\rangle \\
-\left\langle cW_{\rm II}^{(4)}(\tau_{4})W_{\rm I}^{(3)}(\tau_{3})W_{\rm II}^{(2)}(\tau_{2})W_{\rm I}^{(1)}(\tau_{1})\right\rangle
\end{array} 
\right]
\]
Expanding out all possible contractions and implementing the Green functions and noting that
\[
\begin{array}{c@{, \hfill\quad} c}
\left\langle DX(\tau_1)DX(\tau_3) \right\rangle=-2\delta(T) & \left\langle DX(\tau_2)DX(\tau_4)\right\rangle=-2\delta(T) \\
\left\langle DX(\tau_1)DX(\tau_4) \right\rangle=+2\delta(T) & \left\langle DX(\tau_2)DX(\tau_3)\right\rangle=+2\delta(T)                                                                                           \end{array}
\]
With these Green functions in hand we arrive at the following $s$-channel amplitude:
\[
A_{4s} =   \frac{8}{s}\left[ \begin{array}{l}
 +\frac{s}{4}\left( {\epsilon _1  \cdot \epsilon _4 } \right)\left( {\epsilon _2  \cdot \epsilon _3 } \right) -\frac{s}{4}\left( {\epsilon _2  \cdot \epsilon _4 } \right)\left( {\epsilon _1  \cdot \epsilon _3 } \right)-(\frac{s}{4}+ \frac{u}{2})\left( {\epsilon _1  \cdot \epsilon _2 } \right)\left( {\epsilon _4  \cdot \epsilon _3 } \right) \\
  + \left( {\epsilon _2  \cdot k_1 } \right)\left( {\epsilon _4  \cdot k_3 } \right)\left( {\epsilon _1  \cdot \epsilon _3 } \right) + \left( {\epsilon _1  \cdot k_2 } \right)\left( {\epsilon _3  \cdot k_4 } \right)\left( {\epsilon _2  \cdot \epsilon _4 } \right) \\
  + \left( {\epsilon _1  \cdot k_3 } \right)\left( {\epsilon _2  \cdot k_4 } \right)\left( {\epsilon _3  \cdot \epsilon _4 } \right) + \left( {\epsilon _4  \cdot k_2 } \right)\left( {\epsilon _3  \cdot k_1 } \right)\left( {\epsilon _1  \cdot \epsilon _2 } \right) \\
  - \left( {\epsilon _1  \cdot k_2 } \right)\left( {\epsilon _4  \cdot k_3 } \right)\left( {\epsilon _2  \cdot \epsilon _3 } \right) - \left( {\epsilon _3  \cdot k_4 } \right)\left( {\epsilon _2  \cdot k_1 } \right)\left( {\epsilon _1  \cdot \epsilon _4 } \right) \\
  - \left( {\epsilon _1  \cdot k_4 } \right)\left( {\epsilon _2  \cdot k_3 } \right)\left( {\epsilon _3  \cdot \epsilon _4 } \right) - \left( {\epsilon _3  \cdot k_2 } \right)\left( {\epsilon _4  \cdot k_1 } \right)\left( {\epsilon _1  \cdot \epsilon _2 } \right) \\
 \end{array} \right]
\]
A similar calculation can be done for the $t$-channel graph, and the result is simply changing the labeling of all momenta and polarizations in the $s$-channel result according to:
\[\begin{array}{ccc}
    s & \rightarrow & t \\
    1 & \rightarrow & 4 \\
    2 & \rightarrow & 1 \\
    3 & \rightarrow & 2 \\
    4 & \rightarrow & 3
  \end{array}
\]
We arrive at:
\[
A_{4t}  =\frac{8}{t}\left[ \begin{array}{l}
 +\frac{t}{4}\left( {\epsilon _4  \cdot \epsilon _3 } \right)\left( {\epsilon _2  \cdot \epsilon _1 } \right) -\frac{t}{4}\left( {\epsilon _2  \cdot \epsilon _4 } \right)\left( {\epsilon _1  \cdot \epsilon _3 } \right)-(\frac{t}{4}+ \frac{u}{2})\left( {\epsilon _1  \cdot \epsilon _4 } \right)\left( {\epsilon _2  \cdot \epsilon _3 } \right) \\
  + \left( {\epsilon _1  \cdot k_4 } \right)\left( {\epsilon _3  \cdot k_2 } \right)\left( {\epsilon _2  \cdot \epsilon _4 } \right) + \left( {\epsilon _2  \cdot k_3 } \right)\left( {\epsilon _4  \cdot k_1 } \right)\left( {\epsilon _1  \cdot \epsilon _3 } \right) \\
  + \left( {\epsilon _1  \cdot k_3 } \right)\left( {\epsilon _4  \cdot k_2 } \right)\left( {\epsilon _2  \cdot \epsilon _3 } \right) + \left( {\epsilon _2  \cdot k_4 } \right)\left( {\epsilon _3  \cdot k_1 } \right)\left( {\epsilon _1  \cdot \epsilon _4 } \right) \\
  - \left( {\epsilon _1  \cdot k_2 } \right)\left( {\epsilon _4  \cdot k_3 } \right)\left( {\epsilon _2  \cdot \epsilon _3 } \right) - \left( {\epsilon _2  \cdot k_1 } \right)\left( {\epsilon _3  \cdot k_4 } \right)\left( {\epsilon _1  \cdot \epsilon _4 } \right) \\
  - \left( {\epsilon _1  \cdot k_4 } \right)\left( {\epsilon _2  \cdot k_3 } \right)\left( {\epsilon _3  \cdot \epsilon _4 } \right) - \left( {\epsilon _3  \cdot k_2 } \right)\left( {\epsilon _4  \cdot k_1 } \right)\left( {\epsilon _1  \cdot \epsilon _2 } \right) \\
 \end{array} \right]
\]
Adding the two channels again gives the complete 4-point amplitude.

\section{VII. Conclusions}

From the rules of the previous two examples we can extend our approach to higher-points. The fermionic Green functions are constructed as direct products of Green functions of each of the subgraphs that constitute the entire graph.

In this paper, we have derived first-quantized rules for Yang-Mills tree amplitudes. The derivation is based on the BRST quantization of the spinning particle. We derived the vertex operators and vacuum in the BRST formalism, and this ensures background gauge invariance for our amplitude. Rules for tree amplitude calculation have been established after defining the derivative on 1D topological spaces. Such rules should be similar to ones that will be required for higher-loop calculations. It would also be interesting to consider the N=4 spinning particle, where one can couple to background gravity and calculate gravity amplitudes.

\section*{Acknowledgment}

The authors would like to thank the referee for his/her detailed comments. This work was supported in part by National Science Foundation Grant No.\ PHY-0354776.


\begin{thebibliography}{99}

\bibitem{Bern:1991aq}
  Z.~Bern and D.~A.~Kosower,
  Nucl.\ Phys.\ B {\bf 379}, 451 (1992).

\bibitem{Strassler:1992zr}
  M.~J.~Strassler,
  Nucl.\ Phys.\ B {\bf 385}, 145 (1992)
  [arXiv:\phref{9205205}].

\bibitem{Sato:1998sf}
  H.~T.~Sato and M.~G.~Schmidt,
  Nucl.\ Phys.\  B {\bf 560}, 551 (1999)
  [arXiv:\hhref{9812229}].
  
  H.~T.~Sato, M.~G.~Schmidt and C.~Zahlten,
  Nucl.\ Phys.\  B {\bf 579}, 492 (2000)
  [arXiv:\hhref{0003070}].

\bibitem{Brink:1976uf}
  L.~Brink, P.~Di Vecchia and P.~S.~Howe,
  Nucl.\ Phys.\  B {\bf 118}, 76 (1977).

\bibitem{Howe:1988ft}
  F.~A.~Berezin and M.~S.~Marinov,
  Annals Phys.\  {\bf 104}, 336 (1977).

  V.~D.~Gershun and V.~I.~Tkach,
  JETP Lett.\  {\bf 29}, 288 (1979)
  [Pisma Zh.\ Eksp.\ Teor.\ Fiz.\  {\bf 29}, 320 (1979)].


  P.~S.~Howe, S.~Penati, M.~Pernici and P.~K.~Townsend,
  Phys.\ Lett.\  B {\bf 215}, 555 (1988).

  P.~S.~Howe, S.~Penati, M.~Pernici and P.~K.~Townsend,
  Class.\ Quant.\ Grav.\  {\bf 6}, 1125 (1989).

  W.~Siegel,
  Int.\ J.\ Mod.\ Phys.\  A {\bf 4}, 2015 (1989).

  R.~Marnelius and U.~Martensson,
  Nucl.\ Phys.\  B {\bf 321}, 185 (1989).

  N.~Marcus,
  Nucl.\ Phys.\  B {\bf 439}, 583 (1995)
  [arXiv:\hhref{9409175}].

\bibitem{Bastianelli:2002qw}
  F.~Bastianelli, O.~Corradini and A.~Zirotti,
  Phys.\ Rev.\  D {\bf 67}, 104009 (2003)
  [arXiv:\hhref{0211134}].


\bibitem{Siegel:1986zi}
  W.~Siegel and B.~Zwiebach,
  Nucl.\ Phys.\  B {\bf 282}, 125 (1987).

\bibitem{Siegel:1999ew}
  W.~Siegel, {\it Fields}, Chapter XII
  [arXiv:\hhref{9912205}].

\bibitem{Bastianelli:2005vk}
  F.~Bastianelli, P.~Benincasa and S.~Giombi,
  JHEP {\bf 0504}, 010 (2005)
  [arXiv:\hhref{0503155}].

\bibitem{Fields}
  W.~Siegel, {\it Fields}, subsections VIIA2 and VIIIC5
 [arXiv:\hhref{9912205}].

\bibitem{Casalbuoni:1974pj}
  R.~Casalbuoni, J.~Gomis and G.~Longhi,
  Nuovo Cim.\  A {\bf 24}, 249 (1974).

\bibitem{Holzler:2007xt}
  H.~Holzler,
  [arXiv:\nohhref{0704.3392}].
  

\bibitem{Schubert:2001he}
  C.~Schubert,
  Phys.\ Rept.\  {\bf 355}, 73 (2001)
  [arXiv:\hhref{0101036}].

\bibitem{Dai:2006vj}
  P.~Dai and W.~Siegel,
  Nucl.\ Phys.\  B {\bf 770}, 107 (2007)
  [arXiv:\hhref{0608062}].

\end{thebibliography}
\end{document}